\documentclass[12pt,a4]{article}
\usepackage{graphicx}
\usepackage{longtable}
\usepackage{amsmath}
\usepackage[all]{xy}

\newcommand{\be}{\begin{equation}}
\newcommand{\ee}{\end{equation}}
\newcommand{\ba}{\begin{eqnarray}}
\newcommand{\ea}{\end{eqnarray}}
\newcommand{\baa}{\begin{eqnarray*}}
\newcommand{\eaa}{\end{eqnarray*}}

\renewcommand{\k}{{\vec k}}

\def \r{\vec r}
\newcommand{\R}{{\vec R}}
\newcommand{\q}{{\vec q}}

\def \a{\vec a}
\def \K{{\vec K}}

\begin{document}

\title{The complete impurity scattering formalism in graphene}
\author{Cristina Bena$^{1,2}$\\
{\small \it 1 Institute de Physique Th\'eorique, CEA/Saclay, CNRS, URA 2306},
\vspace{-.1in}\\{\small \it  Orme des Merisiers, F-91191 Gif-sur-Yvette, France}
\\{\small \it 2 Laboratoire de Physique des Solides},
\vspace{-.1in}\\{\small \it Universit\'e Paris-Sud, 91405 Orsay CEDEX, France}}
\maketitle

\begin{abstract}
We present the complete formalism that describes scattering in graphene at low-energies. We begin by analyzing the real-space free Green's function matrix, and its analytical expansions at low-energy, carefully incorporating the discrete lattice structure, and arbitrary forms of the atomic-orbital wave function. We then compute the real-space Green's function in the presence of an impurity. We express our results both in $2 \times 2$ and  $4\times 4$ forms (for the two sublattices and the two inequivalent valleys of the first Brillouin zone). We compare this with the $4\times 4$ formalism proposed in Refs.~\cite{falko,glazman}, and show that the latter is incomplete. We describe how it can be adapted to accurately take into account the effects of inter-valley scattering on spatially-varying quantities such as the local density of states.
\end{abstract}

\section{Introduction}

Graphene's most interesting features are the presence of two sublattices and the existence of linearly-dispersing quasiparticles close to the Dirac points, where the valence band and the conduction band touch. The existence of two sublattices (or equivalently of two atoms per unit cell) necessitates the introduction of an extra degree of freedom, often called pseudospin, which gives the Hamiltonian a $2\times 2$ matrix structure. The Hamiltonian matrix can be diagonalized to obtain the band structure, which is linear at low energies.

This allows to extract analytically the low-energy physics by expanding the Hamiltonian close to the nodal points and focusing entirely on the regions in momentum space close to these points.
However, restricting the analysis to low momenta disregards the real-space discrete structure of the lattice. For a monoatomic lattice, the discrete result can be recovered straightforwardly by overlapping the continuous result with the discrete lattice structure. However graphene's two sublattices make the overlap with the lattice less intuitive (a careful analysis of the real-space discrete structure of the Green's function is presently lacking).

Here we derive the first complete real-space Green's function in monolayer graphene that takes into account the discrete structure of the lattice and the presence of the two sublattices. We find that different elements of the matrix Green's function are non-zero on different sublattice sites. We begin by analyzing the full real-space Green's function valid at arbitrary energy. This cannot be evaluated analytically, but can be obtained by performing a two-dimensional numerical integral. Our formalism allows us to describe both localized and extended atomic-orbital wave functions. Subsequently we derive analytically the discrete low-energy real-space Green's function, which we first write as a $2 \times 2$ matrix (using two sublattice indices). This is completely general and incorporates the existence of the two inequivalent valleys without requiring the use of $4 \times 4$ matrices (two sublattice and two valley indices). However, in order to make connection with the previously derived analytical $4 \times 4$ formalism \cite{falko,glazman}, we rewrite the Green's function in  $4 \times 4$ language, and compare our result with the real-space $4 \times 4$ Green's function derived in \cite{glazman}.

We then apply our formalism to describe impurity scattering.
We use the Born approximation to relate the free Green's function to the Green's function in the presence of an impurity, and to calculate the local density of states (LDOS). In the $2 \times 2$ formalism, the density of states can be obtained simply from the trace of the Green's function matrix. However, in the $4\times 4$ formalism the density of states is the sum of the traces of the four blocks that compose the $4\times 4$ matrix Green's function. This modifies the  $4\times 4$ formalism of \cite{falko,glazman}, and allows it to describe correctly the effects of inter-nodal scattering.

Indeed, for localized impurities (in agreement with \cite{bena1,tami}), both the $2\times 2$ and the corrected $4\times 4$ formalisms
describe accurately the existence of the short-wavelength ($R3$) oscillations  \cite{mallet,mallet2} generated by inter-nodal scattering. Moreover, these formalisms predict that these oscillations decay as $1/r$ \cite{bena1},
slower than the long-wavelength $1/r^2$ oscillations generated by intra-nodal scattering \cite{falko,glazman}, and in agreement with recent experimental observations \cite{mallet2}. On the other hand, the original form of the $4 \times 4$ formalism \cite{falko,glazman}, while capturing accurately the features of the intra-nodal Friedel oscillations (FO), cannot capture the short-wavelength inter-nodal FO for any type of impurity.

In section 2 we compute the Green's function at arbitrary energy. In section 3 we analytically evaluate it at low energy: in section 3.1 we approximate the form of the atomic orbitals by delta-functions and we compute the Green's function, writing it as a $2\times 2$ matrix.  In section 3.2 we generalize this to finite-size orbitals. In section 3.3 we rewrite this Green's function in  $4\times 4$ language and compare it to the one proposed in \cite{glazman}. In section 4 we use the Born approximation to compute the generalized Green's function in the presence of an impurity. In section 5 we calculate the dependence of the local density of states (LDOS) on position. We conclude in section 6. In the Appendix we present the derivation of the imaginary-time Green's function in momentum space.

\section{Green's function}

The tight-binding Hamiltonian for a graphene monolayer is given by:
\be
{\cal H}=-t\sum_{\langle i j\rangle}(a_j^{\dagger} b_i+h.c.),
\ee
where $t$ is the nearest-neighbor hopping amplitude, and $\langle i j \rangle $ denotes summing over the nearest neighbors.
In momentum space this Hamiltonian can be written as:
\be
{\cal H}=\int_{\k \in BZ} [a^{\dagger}_{\nu}(\vec{k}) b_{\nu}(\vec{k}) f(\vec{k})+h.c.]
\label{h0}
\ee
where
\be f(\k)= -t ( e^{- i \k \cdot \vec \delta_1} +  e^{- i \k \cdot \vec \delta_2} +  e^{- i \k \cdot \vec \delta_3}) \ \ , \label{f2} \ee and the integral is performed over the first Brillouin zone (BZ). Also, $\vec{\delta}_1\equiv a \sqrt{3} \hat{\bf x}/2+ a \hat{\bf y}/2$, $\vec{\delta}_2\equiv- a \sqrt{3} \hat{\bf x}/2+ a \hat{\bf y}/2$ and $\vec{\delta}_3\equiv-a \hat{\bf y}$ (as depicted in Fig.~1), where $a$ is the distance between two nearest neighbors.
\begin{figure}[htbp]
\begin{center}
\includegraphics[width=4in]{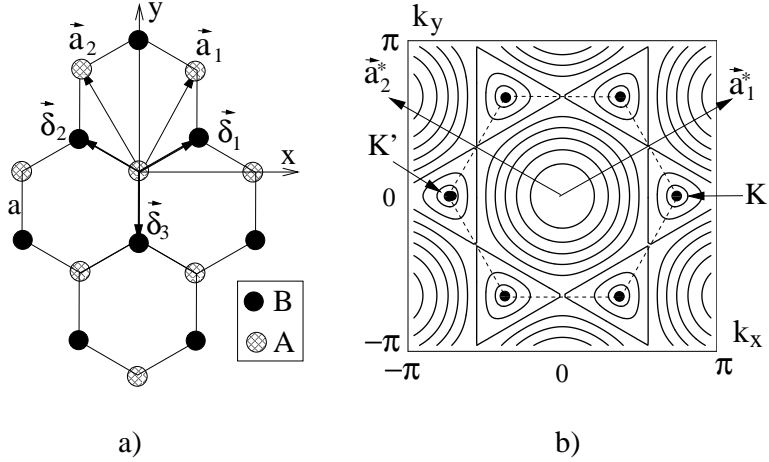}
\vspace{0.15in} \caption{\small Hexagonal honeycomb lattice of graphene (a), and its band structure (b). In b) the equal energy contours are drawn, and the Brillouin zone is indicated by dashed lines. The Dirac points $K$ and $K'$ are marked by arrows, and the reciprocal lattice vectors $\vec{a}^*_{1,2}$ are also drawn.}
\label{fig1}
\end{center}
\end{figure}
As described in \cite{tb}, there are multiple equivalent choices for $f(\k)$, depending on the tight-binding basis used, and on the definition of the Fourier transform (FT) of the real-space operators $a_i$ and $b_i$. The choice (\ref{f2}) (the $II$ basis in Ref.~\cite{tb}) makes the calculations more transparent and will be used throughout this paper.

The real-space imaginary-time matrix Green's function can be defined as:
\be
G_{\alpha \beta}(\vec{r},i \omega_n)\equiv \sum_{\vec{R}_i,\vec{R}_j} \langle c^{ \dagger}_{\alpha j}(i \omega_n) c_{\beta i}(i \omega_n)\rangle \delta(\vec{R}_{\beta j}-\vec{R}_{\alpha i}-\vec{r}),
\label{g0}
\ee
where the sum is performed over all unit cell sites $\vec{R}_{i,j}$, with $\vec{R}_j=n \vec{a}_1+m \vec{a}_2$, and $j=(n,m)$ specifying the position of one graphene unit cell. Here we take $\vec{a}_1= a \sqrt{3} \hat{\bf x}/2+ 3 a \hat{\bf y}/2$, and $\vec{a}_2=- a \sqrt{3} \hat{\bf x}/2+3 a \hat{\bf y}/2 $. Also, $\alpha,\beta=A,B$ such that the operators $c^\dagger_{A i}(i \omega_n)\equiv a^\dagger_i(i \omega_n)$ and $c^\dagger_{B i}(i \omega_n)\equiv b^\dagger_i(i \omega_n)$ denote creating an electron with energy $i \omega_n$ at sites of the sublattice $A$ ($\vec{R}_{A i}=\vec{R}_{i}+\vec{\delta}_A=\vec{R}_i$) and of the sublattice $B$ ($\vec{R}_{B i}=\vec{R}_{i}+\vec{\delta}_B$) respectively, where $\vec \delta_A=0$ and $\vec \delta_B=\vec \delta_3=-\hat{\bf y} a$.

The definition in Eq.~(\ref{g0}) stems from assuming that the atomic orbitals are delta-function localized at the atomic lattice sites. If the atomic orbitals have a finite extent, Eq.~(\ref{g0}) is no longer valid, and the delta-functions need to be replaced by continuous functions. This will be discussed in more detail in section 3.2.

The corresponding momentum-space Green's function can be obtained by simply taking a FT of the real-space Green's function:
\be
G_{\alpha \beta}(\vec{k},i \omega_n)\equiv \int d^2 \vec{r} G(\vec{r},i \omega_n) e^{i \vec{k} \cdot \vec{r}}.
\ee
The advantage of this generalized Green's function is that it is defined for arbitrary values of $\k$, and not only in the first BZ.

We can easily show (see the Appendix for details) that this Green's function can be related to the two-point function of momentum-space operators:
\be
G_{\alpha \beta}(\vec{k},i \omega_n)=\langle c_{\alpha}^{ \dagger}(\k,i \omega_n) c_\beta(\k,i \omega_n)\rangle.
\label{g0m}
\ee
While, by construction, the expectation value is defined only for $\k \in BZ$, as described in the Appendix, the above relation should be understood as evaluating the form of $G(\k)$ inside the BZ, and then expanding the validity of the functional form for arbitrary $\k$.

In order to obtain the retarded real-time unperturbed Green's function we calculate the expectation value with respect to the Hamiltonian (\ref{h0}) using an analytical continuation $i \omega_n\rightarrow \omega+i \delta$:
\be
G^0(\vec{k}, \omega)=\left(%
\begin{array}{cc}
  \omega+i \delta & f(\k) \\
  f^*(\k) & \omega+i \delta \\
\end{array}%
\right)^{-1}.
\ee
By construction, the real-space unperturbed Green's function
can be related to the momentum-space $G^0(\k,\omega)$ by:
\be
G^0(\vec{r},\omega)=\int \frac{d^2 \vec{k}}{4 \pi^2}
G^0(\vec{k},\omega) e^{-i \vec{k} \cdot \vec{r}},
\ee
where we stress again that the integral is performed over the {\it entire} momentum space, and not over the first BZ. Dividing the momentum space in BZ's, we can rewrite the components of the Green's function matrix as:
\ba
G^0_{\alpha \beta}(\vec{r},\omega)&=&\sum_{\vec{Q}_{\mu \nu} \in RL} \int_{\k \in BZ} \frac{d^2 \vec{k}}{4 \pi^2}
G^0_{\alpha \beta}(\vec{k}+\vec{Q}_{\mu \nu},\omega) e^{-i \vec{k} \cdot \vec{r}} e^{-i \vec{Q}_{\mu\nu} \cdot \vec{r}}
\nonumber \\ &=&\sum_{\vec{Q}_{\mu\nu} \in RL} \int_{\k \in BZ} \frac{d^2 \vec{k}}{4 \pi^2}
G^0_{\alpha \beta}(\vec{k},\omega) e^{-i \vec{k} \cdot \vec{r}} e^{-i \vec{Q}_{\mu\nu} \cdot (\vec{r}+\vec{\delta}_{\alpha}-\vec{\delta}_{\beta})}
\ea
where  $\a_1^*= \hat{\bf x} 2 \pi/\sqrt{3} a + \hat{\bf y} 2\pi/3 a $, $\a_2^*=- \hat{\bf x} 2 \pi/\sqrt{3} a + \hat{\bf y} 2\pi/3 a$, such that $\vec{Q}_{\mu\nu}=\mu \vec{a}_1^*+\nu \vec{a}_2^*$ are the reciprocal lattice vectors, and $\mu,\nu$ are arbitrary integers. Here we used the fact that, as described in Ref.~\cite{tb}, $G_{\alpha \beta}^0(\vec{k}+\vec{Q}_{\mu\nu},\omega)=G_{\alpha \beta}^0(\vec{k},\omega) e^{i \vec{Q}_{\mu\nu} \cdot (\vec{\delta}_{\alpha}-\vec{\delta}_{\beta})}$.
The sum over the reciprocal lattice vectors can be performed to yield
\ba
G^0_{\alpha \beta}(\vec{r},\omega)&=&\sum_{j=(m,n)} \delta(\r-\R_j+\vec{\delta}_{\alpha}-\vec{\delta}_{\beta}) \int_{\k \in BZ} \frac{d^2 \vec{k}}{4 \pi^2}
G^0_{\alpha \beta}(\vec{k},\omega) e^{-i \vec{k} \cdot \vec{r}}
\nonumber \\&=&\sum_{j=(m,n)} \delta(\r-\R_j+\vec{\delta}_{\alpha}-\vec{\delta}_{\beta})
\tilde{G}^0_{\alpha \beta}(\vec{r},\omega),
\label{gfull}
\ea
where $\tilde{G}^0(\r,\omega)$ is the reduced real-space Green's function obtained by integrating the momentum-space Green's function only over the first BZ.
We note that the $AA$ and $BB$ components of the Green's function are non-zero only for the sites $R_j$ of the $A$ sublattice (we use the convention that an $A$ atom sits at the origin of the coordinate system), while the $AB$ and $BA$ components are non-zero only for $B$-type sites.

Thus one can obtain the total Green's function by computing the reduced Green's function and multiplying it by the corresponding delta functions in the manner described above.
This is quite general, and valid at arbitrary energy. However the integral (\ref{gfull}) can be evaluated only numerically \cite{bena1,tami,bk,imp2,imp3}.

\section{Low-energy analytical expressions}
\subsection{The Green's function in $2\times2$ language}
At low energy, the Green's function can be obtained in a more elegant way by  observing that only the regions close to the nodal points contribute to the FT integral:
\be
G^0(\vec{r},\omega)=\sum_{\xi=\pm,\eta=(\mu,\nu)}\int \frac{d^2 \vec{q}}{4 \pi^2} G^0_{\xi j}(\vec{q},\omega) e^{-i (\vec{q}+\vec{K}_{\xi \eta}) \cdot \vec{r}}
\label{gsum}
\ee
where \cite{tb}:
$$\K_{\xi \eta}= {\xi} {\a_1^*-\a_2^* \over 3}+ \mu \a_1^* + \nu \a_2^*.$$
Here $\xi=\pm$ is the valley index (there are two such points for each unit cell of the reciprocal space), and $\eta=(\mu,\nu)$, such that $\vec{Q}_{\mu\nu}=\mu\a_1^*+\nu \a_2^* \in RL$.
For example, the most used valley pair containing two corners of the first BZ is described in this notation as
$\K\equiv \K_{+(0,0)}=(4\pi/3\sqrt{3},0)$ and $\K'\equiv \K_{-(0,0)}=(-4\pi/3\sqrt{3},0)$.
Also $G^0_{\xi \eta}(\vec{q})\equiv G^0_{\xi \eta}(\vec{q}+\vec{K}_{\xi \eta})$, and $\vec{q}$ is a small deviation from a nodal point.
Thus we obtain
\be
G^0(\vec{r},\omega)= \sum_{\xi,\eta}G^0_{\xi \eta}(\vec{r},\omega) e^{-i \vec{K}_{\xi \eta} \cdot \vec{r}},
\label{gsum1}
\ee
where
\be
G^0_{\xi \eta}(\vec{r},\omega)\equiv\int \frac{d^2 \vec{q}}{4 \pi^2} G^0_{\xi \eta}(\vec{q},\omega) e^{-i \vec{q} \cdot \vec{r}}.
\ee
The $G^0_{\xi \eta}(\r,\omega)$ functions can be calculated exactly by linearizing $f(\k)$ near the nodal points:
\ba
G^0_{\xi \eta}(\vec{r},\omega)\propto\omega \left(%
\begin{array}{cc}
  H_0^{(1)}({\bf z}) & i z_{\mu \nu} (\xi x-i y) H_1^{(1)}({\bf z})/r \\
i z_{\mu\nu}^*(\xi x+i y) H_1^{(1)}({\bf z})/r  & H_0^{(1)}({\bf z}) \\
\end{array}%
\right),
\ea
where $r=|\r|$, ${\bf z}\equiv |\omega| r/v $, $v=3 ta/2$, $H_{0,1}^{(1)}({\bf z})$ are Hankel functions, and $z_{\mu\nu}=e^{-i \K_{\xi,(\mu,\nu)} \cdot (\vec \delta_{B}-\vec{\delta}_{A})}=e^{2 i \pi (\mu+\nu)/3}$.
Noting that the sum over the reciprocal lattice vectors $(\mu,\nu)$ can be performed to obtain a sum of delta functions centered on the $\R_j$ sites, we obtain:
\begin{small}
\be
G^0(\vec{r},\omega)=\sum_{\xi=\pm} G^0_\xi(\vec{r},\omega),
\label{g4}
\ee
with
\be
G^0_\xi(\vec{r},\omega) \propto \sum_{j=(m,n)}\omega e^{-i \xi \r \cdot \vec{K}}\left(%
\begin{array}{cc}
\delta(\r-\R_j)  H_0^{(1)}({\bf z}) & i \delta(\r-\R_j-\vec{\delta}_B)  (\xi x-i y) H_1^{(1)}({\bf z})/r \\
i \delta(\r-\R_j+\vec{\delta}_B)(\xi x+i y) H_1^{(1)}({\bf z})/r  & \delta(\r-\R_j) H_0^{(1)}({\bf z}) \\
\end{array}%
\right)
\label{gxi}
\ee
\end{small}

\noindent This can be evaluated to yield
\begin{small}
\ba
G^0(\vec{r},\omega)\propto 2\sum_{j=(m,n)} \omega
\left(%
\begin{array}{cc}
\delta(\r-\R_j)  H_0^{(1)}({\bf z}) \cos(\vec{K} \cdot \r)&  \delta(\r-\R_j-\vec{\delta}_B)  \phi_{1}(\r) H_1^{(1)}({\bf z}) \\
\delta(\r-\R_j+\vec{\delta}_B)\phi_{2}(\r)H_1^{(1)}({\bf z})  & \delta(\r-\R_j) \cos(\vec{K} \cdot \r)H_0^{(1)}({\bf z}) \\
\end{array}%
\right)
\label{glow}
\ea
\end{small}

\noindent where $\phi_{1,2}(\r)=[\sin(\vec{K} \cdot \r) x\pm\cos(\vec{K} \cdot \r) y]/r$, $\vec{K}=\vec{K}_{+(0,0)}=(\vec{a}_1^*-\vec{a}_2^*)/3$. The factors $e^{i \vec{K} \cdot \r}$ in the Green's function have also been discussed in Ref.~\cite{basko}.

\subsection{Extended atomic orbitals}

The above derivation of the Green's function is valid if the atomic orbitals are fully localized. More generally, we can take into account the finite extension of the atomic orbitals by replacing the delta-functions in the definition (\ref{g0}) of the Green's function by a finite-extent function $\phi(\r)$, centered at $\r=0$:
\be
G_{\alpha \beta}(\vec{r},i \omega_n)=\sum_{\vec{R}_i,\vec{R}_j} \langle c^{ \dagger}_{\alpha j}(i \omega_n) c_{\beta i}(i \omega_n)\rangle \phi(\vec{R}_{\beta j}-\vec{R}_{\alpha i}-\vec{r}).
\label{g0e}
\ee
A possible choice for $\phi$ is a Gaussian ($\phi(\r)=e^{-r^2/l_a^2}$), where $l_a$ is the ``radius'' of an atomic orbital.
We can see from the derivation in Eq.~(\ref{app1}) of the Appendix, that the FT of the real-space Green's function is no longer equal to a two-point function of momentum-space operators. For simple forms of $\phi$ such as the Gaussian, the integrals in Eq.~(\ref{app1}) can still be performed analytically, but the two-point function  in Eq.~(\ref{app2}) should be multiplied by a momentum-dependent factor that decays at large momenta.
In the low-energy limit, this translates into a decay of $G_{\xi \eta}^0(\q,\omega)$ with the magnitude of $\K_{\xi \eta}$. Hence the nodal points do not contribute equally to the sum in Eq.~(\ref{gsum1}), which cannot in general be evaluated analytically. Therefore the analytic form (\ref{gxi},\ref{glow}) of the Green's function can only be used when the extent of the atomic orbitals is negligible \footnote{If the orbitals are not fully localized, the Green's function in Eqs.~(\ref{gxi},\ref{glow}) also breaks the hexagonal symmetry of the lattice.}.

If the extent of the atomic orbitals is finite, the Green's function can be obtained from Eq.~(\ref{g0e}). The two-point function of the $c_{\alpha,\beta}$ operators is simply the Fourier transform of the momentum-space two-point function (with the momentum integral performed over the first BZ), which needs to be overlapped with the atomic-orbital wave-functions using Eq.~\ref{g0e}.

However, this integral over the first BZ cannot be evaluated analytically.
To obtain the low-energy analytical behavior of the Green's function one can start from Eq.~(\ref{gsum1}) and use the fact that, for finite-size atomic orbitals, $G_{\xi \eta}^0(\q,\omega)$ decays with the magnitude of $\K_{\xi \eta}$. This implies that the sum in Eq.~(\ref{gsum1}) is dominated by the nodal points that are closest to the origin of the first BZ. Their exact number, and corresponding weights depend on the exact atomic-orbital function. The simplest approximation is to consider that only the nodal points at the corners of the first BZ contribute \cite{bena1}. This yields:
\be
G^0(\vec{r},\omega)=\sum_{\xi=\pm}G^0_\xi(\vec{r},\omega),
\ee
with
\be
G^0_\xi(\vec{r},\omega) \propto\omega \left(%
\begin{array}{cc}
{\vec {\bf \Phi}}_\xi(\r) \cdot {\vec {\bf I}} H_0^{(1)}({\bf z}) & i {\vec {\bf \Phi}}_\xi(\r) \cdot { \vec {\bf Z}}_\xi (\xi x-i y) H_1^{(1)}({\bf z})/r \\
i {\vec {\bf \Phi}}_\xi(\r) \cdot {\vec {\bf Z}_\xi^*}(\xi x+i y) H_1^{(1)}({\bf z})/r  & {\vec {\bf \Phi}}_\xi(\r) \cdot {\vec {\bf I}} H_0^{(1)}({\bf z}) \\
\end{array}%
\right),
\label{gxi2}
\ee
where ${\vec{\bf \Phi}}_\xi(\r)=(e^{-i \r \cdot \vec{K}_{\xi (0,0)}},e^{-i \r \cdot \vec{K}_{\xi (0,\xi)}},e^{-i \r \cdot \vec{K}_{\xi (-\xi,0)}})$, ${\vec{\bf I}}=(1,1,1)$  and ${\vec{\bf Z}}_\xi=(1,e^{2 i\pi\xi/3},e^{-2 i \pi \xi/3})$.

\subsection{The low-energy analytical form for the Green's function in the $4\times4$ basis}

It is interesting to note that, instead of performing the sum over $\xi$, one can also write the Green's function as a $4\times 4$ matrix, with indices corresponding to the $A_+,B_+,A_-,B_-$ components:
\be
G^0_{\xi \xi'}(\vec{r},\omega)=\delta_{\xi\xi'} G^0_{\xi}(\vec{r},\omega).
\label{gxi4}
\ee
We can then compare the above Green's function (where $G^0_{\xi}(\vec{r},\omega)$ was derived in Eq.~(\ref{gxi})), to the $4\times4$ Green's function proposed in Ref.~\cite{glazman}. The latter has a similar structure, with $G^0_{\xi}(\vec{r},\omega)$ being given by:
\be
G^0_\xi(\vec{r},\omega) \propto \left(%
\begin{array}{cc} H_0^{(1)}({\bf z}) & i  (\xi x-i y) H_1^{(1)}({\bf z})/r \\
i (\xi x+i y) H_1^{(1)}({\bf z})/r  &  H_0^{(1)}({\bf z}) \\
\end{array}%
\right)
\label{gxi-glazman}
\ee
The main differences between this Green's function and the Green's function we derived in (\ref{gxi}) for localized orbitals are 1) the discrete structure, and 2) the presence of the oscillatory terms $e^{i \K\cdot \r}$. These two factors are crucial to explain the short-scale fluctuations of the system.

\section{The generalized Green's function in the presence of an impurity}

To describe systems that are not translationally invariant (like in the presence of an impurity), one needs the ``full" Green's function matrix, defined as
\be
{\cal G}_{\alpha \beta}(\vec{r}_1,\r_2,i \omega_n)\equiv \sum_{\vec{R}_i,\vec{R}_j} \langle c^{ \dagger}_{\alpha j}(i \omega_n) c_{\beta i}(i \omega_n)\rangle \delta(\vec{R}_{\alpha i}-\vec{r}_1) \delta(\R_{\beta j}-\r_2).
\label{ggd}
\ee

For a translationally invariant system with one atom per unit cell, the relation between the usual Green's function and the ``full'' one is simply $G^0(\r_2-\r_1,i \omega_n)={\cal G}^0(r_1,r_2,i\omega_n)$. Indeed, for a monoatomic system, the difference $\r_1-\r_2$ can be a Bravais lattice vector only if $\r_1$ and $\r_2$ are both Bravais lattice vectors. This yields\footnote{We use a normalization of the delta function in which $\delta(\r-\R_i)=N$ if $\r=\R_i$ and $0$ otherwise, where $N$ is the total number of unit cells in the crystal}: $\sum_{i,j} \delta(\r_1-\r_2-\R_i+\R_j)= \sum_{i,j} \delta(\r_1-\R_i)\delta(\r_2-\R_j)$. For graphene, things are however a bit more complicated. Thus, if $\r_2$ is on a site of type $A$, the components ${\cal{G}}^0_{\alpha B}$ are zero, while the components $G^0_{\alpha B}$ are not. In general the relation between ${\cal G}^0$ and $G^0$ can be written as
\be
{\cal{G}}^0_{\alpha \beta}(\r_1,\r_2,i \omega_n)=\sum_{j=(m,n)} G^0_{\alpha\beta}(\r_2-\r_1) \delta(\r_2-\R_{\beta j}).
\label{gg}
\ee

We would now like to use the above formalism to carefully calculate the corrections to the Green's functions in the presence of an impurity. In the Born approximation\footnote{Our analysis also applies to the T-matrix approximation if the impurity potential is localized.},
the impurity corrections to the ``full'' Green's function are given by:
\be
\Delta {\cal{G}}(\r_1,\r_2,i \omega_n)=\int_{\r_3} {\cal{G}}^0(\r_1,\r_3,i \omega_n)\cdot {\hat u}(\r_3) \cdot {\cal{G}}^0(\r_3,r_1,i \omega_n),
\ee
where $\hat u(\r)$ is the scattering potential matrix, and the dot ('$\cdot$') denotes matrix multiplication.

At low energy, an analytical expression for ${\cal{G}}^0$ can be obtained directly from Eq.~(\ref{glow}) and substituted in the above expression. However, it is sometimes more transparent to use the expansion (\ref{gsum}) to write:
\be
\Delta {\cal{G}}(\r_1,\r_2,i \omega_n)=\sum_{\xi,\xi'\eta,\eta'} \int_{\r_3} {\cal{G}}^0_{\xi \eta}(\r_1,\r_3,i \omega_n) \cdot \hat u(\r_3) \cdot {\cal{G}}^0_{\xi' \eta'}(\r_3,\vec{r}_2,i \omega_n) e^{i(\vec{K}_{\xi \eta}\cdot \r_1-\vec{K}_{\xi' \eta'}\cdot \vec{r}_2)}
\label{gsum3}
\ee
For $\r_1=\r_2$ this equation allows us to see explicitly that each inter-nodal scattering process contributes at a wavevector corresponding to the difference between $\vec{K}_{\xi j}$ and $\vec{K}_{\xi' j'}$. Also, we can easily modify Eq.~(\ref{gsum3}) to approximate the spatially dependent matrix $\hat u(\r_3)$ by a matrix $\hat u$ which is independent of position, but that depends explicitly on $\xi,\xi',\eta$ and $\eta'$:
$\hat u(\r_3) \rightarrow \hat u_{\xi \xi'}^{\eta \eta'}$\footnote{For example, for a localized impurity, the scattering is independent of momentum, so $\hat u$ should be independent of $\xi,\xi',\eta$ and $\eta'$. However, if the impurity is extended, the inter-nodal scattering is suppressed and $\hat u_{\xi \xi'}^{\eta \eta'} \propto \delta_{\xi \xi'} \delta{\eta \eta'}$.}.

In order to make connection with the $4\times4$ formalism described in \cite{falko,glazman} we rewrite (\ref{gsum3}) as
\be
\Delta {\cal{G}}(\r_1,\r_2,i \omega_n)=\sum_{\xi,\xi'} \Delta {\cal{G}}_{\xi\xi'}(\r_1,\r_2,i \omega_n)
\label{xi}
\ee
with
\be
\Delta{\cal{G}}_{\xi\xi'}(\r_1,\r_2,i \omega_n)=\int_{\r_3} {\cal{G}}^0_{\xi}(\r_1,\r_3,i \omega_n) \cdot \hat u(\r_3) \cdot {\cal{G}}^0_{\xi'}(\r_3,\vec{r}_2,i \omega_n)
\ee
where ${\cal{G}}^0_\xi$ can be extracted from  Eqs.~(\ref{gxi},\ref{gg}).
This corresponds to a $4\times4$ expression:
\be
\Delta {\cal{G}}_{\xi\xi'}(\r_1,\r_2,i \omega_n)=\int_{\r_3} \sum_{\xi_1,\xi_2}{\cal{G}}^0_{\xi\xi_1}(\r_1,\r_3,i \omega_n) \cdot \hat u_{\xi_1 \xi_2}(\r_3) \cdot {\cal{G}}^0_{\xi_2\xi'}(\r_3,\vec{r}_2,i \omega_n),
\ee
where ${\cal{G}}^0_{\xi_1\xi_2}\equiv \delta_{\xi_1\xi_2} {\cal{G}}^0_{\xi_1}$ (\ref{gxi2}).

However, as can be seen from Eq.~(\ref{xi}), the physical properties of graphene are not related directly to the $4 \times 4$ Green's function matrix, but to the sum of the four $2\times2$ blocks that compose it. Thus, the density of states is given by
\be
\rho(\r,\omega)=-\frac{1}{\pi}\sum_{\xi,\xi'}{\rm Im} \{{\rm Tr}[ {\cal G}_{\xi\xi'}(\r,\r,i\omega_n \rightarrow \omega+i \delta)]\}.
\label{ldos}
\ee
This differs from
$$\rho(\r,\omega)=-\frac{1}{\pi}\sum_{\xi}{\rm Im} \{{\rm Tr}[ {\cal G}_{\xi\xi}(\r,\r,i\omega_n \rightarrow \omega+i \delta)]\},$$
which was proposed in \cite{falko,glazman}.  Together with the form of the full Green's function presented in Eq.~(\ref{glow}), our prescription (\ref{ldos}) is a crucial factor in describing correctly the effects of inter-nodal $\xi\rightarrow\xi'$ scattering using a $4\times4$ formalism.

\section{The LDOS in the presence of a localized impurity}

We now focus on a simple limit when the impurity is localized on an $A$ atom, (delta-function potential), for which we obtain:

\be
\Delta {\cal{G}}(\r_1,\r_2,i \omega_n)={\cal{G}}^0(\r_1,0,i \omega_n)\cdot \hat u \cdot {\cal{G}}^0(0,\vec{r_2},i \omega_n)
\ee
where $\hat{u}=\{\{1,0\},\{0,0\}\}$. For localized atomic orbitals this yields:
\ba
\rho(\r)&=&-\frac{1}{\pi} {\rm Im}[G_{AA}^0(-\r,\omega) G_{AA}^0(\vec{r},\omega)+G_{AB}^0(-\r,\omega) G_{BA}^0(\vec{r},\omega)]
\nonumber \\&\propto&
-\frac{1}{\pi} \sum_{j=(m,n)}{\rm Im}\big[\delta(\r-\R_{Aj}) H_0^{(1)}({\bf z})^2 \cos(\K\cdot\r)^2+\delta(\r-\R_{Bj})\phi_1(\r)^2 H_1^{(1)}({\bf z})^2\big]
\ea
Far away from the impurity (${\bf z} \gg 1$) the impurity gives rise to $1/r$ decaying oscillations on both sublattices. As mentioned in \cite{bena1,tami,sciencef}, the oscillations due to intra-nodal scattering have alternating signs on the $A$ and the $B$ sublattices.
Thus the dominant $1/r$ behavior cancels if one coarse-grains the system so that the $A$ and $B$ atoms in a unit cell are measured together. This is what is expected in a real-space LDOS measurement in a real graphene sample, due to the fact that the $A$ and the $B$ atomic orbitals have a finite overlap.
The $A$ and the $B$ oscillations are also mixed together when one measures the FT of the local density of states. Indeed, the intra-nodal oscillations give rise to a disc in the center of the BZ corresponding to $1/r^2$ decaying oscillations, and not to a ring-like singularity, as for $1/r$ oscillations \cite{bena1}.

For inter-nodal scattering, we can see that the $1/r$ decaying oscillations do not cancel between two neighboring $A$ and $B$ atoms \cite{bena1,tami}. This gives rise to much more robust, short-wavelength oscillations, which can be observed for larger areas around the impurity. These translate into rings of high intensity in the FT of the LDOS close to the corners of the BZ. The difference between the intra-nodal and the inter-nodal FO was confirmed by a recent experiment \cite{mallet2}, which observed no ring in monolayer graphene close to the center of the BZ, and detected rings close to the corners of the BZ.

We should note that by using the formalism in Refs.~\cite{falko,glazman}, the $1/r$ decaying short-wavelength oscillations are not accurately retrieved. For example, no $1/r$ decaying inter-nodal oscillations arise, even in the presence of a localized impurity. As we show here, this can be corrected by using the expression of the Green's function presented in Eq.~(\ref{glow}), and the correct prescription (\ref{ldos}) to extract the LDOS starting from the $4\times4$ ``full'' Green's function.



\section{Conclusions}

We calculated the complete low-energy free Green's function for monolayer graphene. We expanded carefully the Green's function at low energy, paying particular attention to incorporate the effects of the discrete structure of the graphene bipartite lattice, and of the form of the atomic-orbital wave function. We have also calculated the ``full'' Green's function in the presence of impurity scattering. We have found that the $4\times4$ formalism proposed in \cite{falko,glazman} to study the low-energy physics of graphene, while describing correctly intra-nodal scattering, does not accurately capture the effects of inter-nodal scattering on local quantities, such as the LDOS. We have shown that the missing pieces in the  $4\times4$ formalism in \cite{falko,glazman} are 1) the discrete structure 2) the presence of the oscillatory terms $e^{i \K\cdot \r}$ in the Green's function and 3) the prescription to compute the physical properties not from the $4\times4$ matrix itself, but rather from the sum of the four $2\times2$ block that make up the $4\times4$ matrix. We believe it is important to check whether, besides the LDOS, there exist other spatially-dependent quantities that are inaccurately described by the incomplete $4\times 4$ formalism. We have also presented a $2\times 2$ formalism that is more general, and can be used straightforwardly for any type of disorder calculation, and for any type of atomic-orbital wave function. It would be interesting to use our formalism to calculate the real-space dependence of the LDOS in the presence of an extended impurity.
\vspace{.2in}\\
{\noindent \bf \Large Acknowledgements}
\vspace{.2in}\\
We would like to thank Gilles Montambaux, Vladimir Falko, Leonid Glazman, Felix von Oppen, Konstantin Efetov and Denis Ullmo for useful discussions and insightful comments. This work was supported by a Marie Curie Action under the Sixth Framework Programme.

\vspace{.2in}
{\noindent \bf \Large Appendix}
\vspace{.2in}\\
Here we show that the FT of the continuum real-space Green's function defined in Eq.~(\ref{g0}) corresponds indeed to the two-point function of momentum space operators:
\be
G_{\alpha \beta}(\vec{r},i \omega_n)=\sum_{\vec{R}_i,\vec{R}_j} \langle c^{ \dagger}_{\alpha j}(i \omega_n) c_{\beta i}(i \omega_n)\rangle \delta(\vec{R}_{\beta j}-\vec{R}_{\alpha i}-\vec{r})~.
\ee
The FT of the real-space Green's function is
\ba
G_{\alpha \beta}(\vec{k},i \omega_n)&\equiv& \int d^2 \vec{r} G_{\alpha \beta}(\vec{r},i \omega_n) e^{i \vec{k} \cdot \vec{r}} \nonumber \\
&=&\int d^2 \vec{r} \sum_{\vec{R}_i,\vec{R}_j} \langle c^{ \dagger}_{\alpha j}(i \omega_n) c_{\beta i}(i \omega_n)\rangle \delta(\vec{R}_{\beta i}-\vec{R}_{\alpha j}-\vec{r})e^{i \vec{k} \cdot \vec{r}}
\nonumber \\
&=&\sum_{\vec{R}_i,\vec{R}_j} \int_{\k_1 \in BZ} \int_{\k_2 \in BZ} \langle c^{ \dagger}_{\alpha}(\k_1,i \omega_n) c_{\beta}(\k_2,i \omega_n)\rangle e^{i (\k_1-\k)\cdot \vec{R}_{\alpha i}} e^{i (\k-\k_2)\cdot \vec{R}_{\beta j}}
\label{app1}
\ea
Noting that $\sum_{\vec{R}_i}  e^{i (\k-\k')\cdot \vec{R}_{\alpha i}}=\sum_{\vec{Q}_{\mu\nu}\in RL} \exp(-i \vec{Q}_{\mu\nu} \cdot \vec{\delta}_{\alpha}) \delta(\k-\k'+\vec{Q}_{\mu\nu})$, where $\vec{Q}_{\mu\nu}=\mu \a_1^*+\nu \a_2^*$ is any vector of the reciprocal lattice, $\vec{\delta}_A=0$, and $\vec{\delta}_B=(0,-a \hat{\bf y})$, we have
\ba
G_{\alpha \beta}(\vec{k},i \omega_n)=&&\langle c^{ \dagger}_{\alpha}(\k+\vec{Q}_{\mu\nu},i \omega_n) c_{\beta}(\k+\vec{Q}_{\mu\nu},i \omega_n)\rangle|_{\k+\vec{Q}_{\mu\nu}\in BZ} \times
\nonumber \\&& \times \exp[-i \vec{Q}_{\mu\nu} \cdot (\vec{\delta}_{\alpha}-\vec{\delta}_{\beta})]~.
\ea
We can see that this relation holds for arbitrary $k$, inside and outside the first BZ. Using the fact that $c_{\beta}(\k+\vec{Q}_{\mu\nu},i \omega_n)=c_{\beta}(\k) e^{-i \vec{Q}_{\mu\nu}\cdot \vec{\delta}_{\beta}}$, and $c_{\alpha}^\dagger(\k+\vec{Q}_{\mu\nu},i \omega_n)=c_{\alpha}(\k) e^{i \vec{Q}_{\mu\nu}\cdot \vec{\delta}_{\alpha}}$ we obtain
\be
G_{\alpha \beta}(\vec{k},i \omega_n)=\langle c_{\alpha}^{ \dagger}(\k,i \omega_n) c_\beta(\k,i \omega_n)\rangle~.
\label{app2}
\ee
We should note that if one uses the alternative (I) basis presented in Ref.~\cite{tb}, the different components of the matrix Green's function acquire different phases, consistent with the fact that the density of states written in this basis in momentum space is a superposition of two-point functions of A and B operators with a relative phase factor \cite{tb}.

Another observation one should make is that if we repeat the above exercise for the ``full'' Green's function defined in Eq.~(\ref{ggd}), we find:
\ba
{\cal{G}}(\k_1,\k_2,i \omega_n)&\equiv& \int d^2 \vec{r}_1\int d^2 \vec{r}_2 {\cal{G}}(\r_1,\r_2) e^{i \vec{k}_1 \cdot \vec{r}_1}e^{i \vec{k}_2 \cdot \vec{r}_2}
\nonumber \\
&=& \langle c_{\alpha}^{ \dagger}(\k_1,i \omega_n) c_\beta(\k_2,i \omega_n)\rangle~.
\ea

\end{document}